\documentclass[prd,superscriptaddress,altaffilletter,showpacs,nofootinbib]{revtex4-1}
\usepackage{amssymb}
\usepackage{epsfig}
\usepackage{graphicx}
\usepackage{float}
\usepackage{color}
\usepackage{epstopdf}
\usepackage[colorlinks=true,linkcolor=blue,urlcolor=blue]{hyperref}

\usepackage{indentfirst}
\newcommand{\be}{\begin{equation}}
\newcommand{\ee}{\end{equation}}
\newcommand{\bea}{\begin{eqnarray}}
\newcommand{\eea}{\end{eqnarray}}

\newcommand{\vphi}{\varphi}

\begin{document}
 \title[Fluctuations in a thick braneworld with non--minimal coupling  and Gauss--Bonnet term ]{Study
 of field fluctuations and their localization in 
 a thick braneworld generated by gravity non--
 minimally coupled to a scalar field with the  Gauss--Bonnet term}
 
\author{Gabriel Germ\'an}\email{gabriel@fis.unam.mx}
\affiliation{Instituto de Ciencias F\'{\i}sicas, Universidad
Nacional Aut\'onoma de M\'exico.\\
Apdo. Postal 48--3, 62251, Cuernavaca, Morelos, M\'{e}xico.}

\author{Alfredo Herrera--Aguilar}\email{aha@fis.unam.mx}
\affiliation{Instituto de Ciencias F\'{\i}sicas, Universidad
Nacional Aut\'onoma de M\'exico.\\
Apdo. Postal 48--3, 62251, Cuernavaca, Morelos, M\'{e}xico.}
\affiliation{Mesoamerican Centre for Theoretical Physics,
Universidad Aut\'onoma de Chiapas.\\ Km. 4 Carretera Emiliano 
Zapata, CP 29000, Tuxtla Guti\'errez, Chiapas, M\'{e}xico.}
\affiliation{Instituto de F\'{\i}sica y Matem\'{a}ticas,
Universidad Michoacana de San Nicol\'as de Hidalgo.\\
Edificio C--3, Ciudad Universitaria, CP 58040, Morelia, Michoac\'{a}n, M\'{e}xico.}

\author{Dagoberto Malag\'on--Morej\'on}\email{malagon@fis.unam.mx}
\affiliation{Instituto de Ciencias F\'{\i}sicas, Universidad
Nacional Aut\'onoma de M\'exico.\\
Apdo. Postal 48--3, 62251, Cuernavaca, Morelos, M\'{e}xico.}

\author{Israel Quiros}\email{iquiros@fisica.ugto.mx}
\affiliation{Departamento de Matem\'aticas, Centro Universitario de Ciencias Exactas
e Ingenier\'{\i}as.\\ Corregidora 500 S.R.,
Universidad de Guadalajara, 44420 Guadalajara, Jalisco, M\'exico.}

\author{Rold\~ao da Rocha}\email{roldao.rocha@ufabc.edu.br}
\affiliation{Centro de
Matem\'atica, Computa\c c\~ao e Cogni\c c\~ao, Universidade Federal do ABC.\\
Rua Santa Ad\'elia, 166 09210-170, Santo Andr\'e, SP, Brazil.}

\date{\today}
\begin{abstract}
In this work we study a scenario with a warped 5D smooth braneworld with 4D
Minkowski geometry builded from bulk scalar  matter non--minimally coupled to gravity with an additional Gauss--Bonnet term.
We present exact solutions for the full braneworld configuration in contrast to previous
results where only approximate solutions were constructed due to the highly non--linear character
of the relevant differential equations. These solutions allow us to study the necessary
conditions for the finiteness of the 4D Planck mass and additionally,  enables us  to perform
a more rigorous analysis of 4D gravity localization compared to approximate approaches.
It is remarkable that all the constructed braneworld configurations lead to standard 4D gravity localization
since they contain a localized massless tensor mode (the graviton).
We also analyze the localization properties of scalar, vector and
tensor fluctuation modes for the constructed field configurations. We show that for the considered
backgrounds, only the massless tensor mode, i.e. the 4D graviton, is localized on the brane,
while the vector and scalar modes are not confined to the brane.
\end{abstract}
\maketitle

\section{Introduction}
Braneworld models \cite{akama}--\cite{rs2} are an
interesting alternative to the standard  Kaluza--Klein compactification. In order to protect the standard 4D
physics of some unobserved effects, in the Kaluza--Klein paradigm extra dimensions
are compactified to a tiny size (for a 4D observer this is equivalent to a very high energy regime). The braneworld scenario
has another approach, in this alternative the  physics of our 4D world is
compatible with the existence of  infinite extra dimensions. A procedure to do that is by considering that
the Standard Model (SM) fields are  trapped
on a 4D hypersurface, called a 3-brane (our universe).
In contrast with ordinary matter, gravity and exotic matter
can reside in the whole higher-dimensional manifold (bulk).
Although the gravitational field can propagate through all dimensions,  one of the
first requirements to have realistic braneworlds models is to recover standard 4D gravity on the brane.
For example,  if the induced  geometry on the brane is flat we need to localize a 4D graviton on our brane.

The braneworld models are classified with respect to their width in thin and thick branes.
In scenarios with a thin braneworld
\cite{merab, rs, rs2} the curvature scalars are singular at the location of the branes, that is a
direct consequence of the null width of the branes. In spite of that
complication, the so--called Israel-Lanczos junction
conditions \cite{Israel,Shiromizu,Maartens} make harmless such singularities when we study the localization of 
gravity and matter as well as when computing several effective parameters of the model,
as for instance the 4D Planck mass and the 4D cosmological constant.

The fact of considering that ordinary matter fields are completely confined on a
region with  null width is just an approximation.
The phenomenology of the Standard Model is well known at electro--weak
energy scale $m_{EW}$, therefore, although
the standard matter  can not move freely through the extra
dimensions, it is possible that they can access to
distances $r \leq m^{-1}_{EW} \sim 10^{-19}$ m  without
contradictions  with the Standard Model \cite{arkani}.
In other words, there are  more  realistic
alternatives to  thin brane configurations; they are known,
generically, as thick braneworlds or domain walls.
When the brane has a non--trivial width, there is no Israel-Lanczos junction
condition,  and this  might be a mathematical advantage when the
action of the model is complicated (when it contains, for instance, too many matter fields, non--minimal couplings, higher order curvature terms, etc).

Thick branes might be constructed by using different procedures.
One of the most evident examples consists in replacing the delta functions in the action by
non--singular source functions. Another simple method is  by using self-interacting
scalar fields minimally  coupled to
gravity as done in \cite{gremm}--\cite{liuetal}.
A more elaborated procedure to generate a thick brane configuration is by introducing a non-minimal coupling
between gravity and matter.
This kind of interactions arises in several contexts, e.g., it  appears naturally
in cosmology, the Brans--Dicke theory, supergravity and in all the known effective low
energy string theory models (see \cite{FujiiMaeda} and references therein). Within the framework of {\it thin
braneworlds}, the Randall-Sundrum scenario has been modified by considering bulk self-interacting scalar fields non--minimally coupled to gravity, for instance, in \cite{farakos}--\cite{Ahmed}. Thick brane generalizations 
of this kind of models have been considered in \cite{andrianov,liuNM}.

The perturbative stability of these non--minimally coupled brane configurations was explored in Refs. \cite{farakos2,andrianov}. Namely, in \cite{farakos2} the authors performed a stability analysis of a perturbed trivial scalar field and obtained instability regions characterized by certain value of the non--minimal coupling parameter in the conformal limit. Furthermore, in \cite{andrianov} it was shown that when the scalar field is non--trivial, the instability region completely disappears under linear perturbations for any value of the non--minimal coupling parameter. Later on localization of gravity and various matter fields was considered in this kind of braneworld configurations in \cite{liuNM}, where deformed branes were also obtained.

On the other hand, it is well known that the standard Einstein--Hilbert action can be supplemented
by higher order curvature corrections without
generating, in the equations of
motion, terms containing three or higher order derivatives of the metric with respect to the space-time
coordinates \cite{lovelock-madore}. The particular combination which satisfies this requirement
is known as Gauss-Bonnet (GB) invariant. Although in 4D this term has a topological origin which
does not contribute to the
classical equations of motion\footnote{In scenarios with $AdS_{4}$ the 4D Gauss--Bonnet
term produces non--trivial
results in the conserved quantities of the theory \cite{aros}.}, in dimensions higher
than four the GB invariant has a non trivial contribution to the
dynamical equations \cite{madore1,madore2}. In our work the
space-time has five dimensions, on it the  Gauss-Bonnet term takes the following form
\be
{\cal R}^2_{\rm GB}=R^{ABCD} R_{ABCD}-4R^{AB} R_{AB}+R^2,
\label{GB-term}
\ee
where indices in capital roman letters $A,B,C,D$ range over the five bulk dimensions.

Another property  of the  GB combination in dimensions higher than
four is that it leads to a ghost-free theory, furthermore
this invariant is present in different higher dimensional models. For instance, in string theory the GB-term
appears in the first string tension correction to the (tree-level) effective action
\cite{GB-string1}--\cite{GB-string7}.

In the context of braneworld scenarios, the influence of higher curvature terms in
scalar-field-generated thick brane models has been studied in Refs. \cite{mg00}--\cite{Hideki}.
In particular, within the framework of thick braneworld scenarios generated by a self-interacting
minimally coupled scalar field, the GB invariant has been studied in connection with the
localization properties of the various modes of the geometry in Ref. \cite{mg00} (see also, for
instance, \cite{corradini}). A next step towards modifying the above scalar-tensor models of thick braneworlds, would be to assume a non-minimal coupling of the scalar field with the curvature Ricci scalar \cite{HengG,HengG1,HengG2}, which means that the higher-dimensional gravitational coupling is point-dependent. In other words, the gravitational interactions in the bulk are jointly propagated by the higher--dimensional graviton and by the scalar field. The latter drives the strength of gravity by giving a local dynamics to the gravitational coupling. In 4D this kind of coupling is motivated from requiring compatibility with the Mach principle \cite{brans}.

In this work we develop further previous results obtained in
\cite{CQG29} where a 5D thick braneworld is modeled by a smooth scalar domain wall non--minimally coupled to gravity with a Gauss-Bonnet term on the bulk.
The field equations that describe the thick braneworld dynamics are difficult when the non--minimal coupling and the Gauss--Bonnet term are turned on
simultaneously. For this reason, in the articles \cite{pasipoularides} and \cite{CQG29} the authors were able to obtain just
approximate solutions to the field equations
when both effects are present.
Therefore, one of the purposes of our research is to obtain exact solutions for a 5D thick braneworld configuration which accounts for both
a non-minimal coupling of the scalar field to gravity and the Gauss--Bonnet invariant on the bulk.

On the other hand, the geometry of the braneworld models we shall consider possesses a broken 5D Poincar\'e symmetry, but 
preserves the 4D Poincar\'e one. As a consequence of both these facts it follows that the fluctuations of the system can be classified 
into tensor, vector and scalar sectors with respect to the symmetry group $SO(3,1)$, and all of these sectors possess non-trivial dynamics that, in principle, can influence the 4D phenomenology of the system.

A careful analysis of the localization of tensor, vector and scalar fluctuations of 
the fields that generate a braneworld configuration needs to have exact solutions at hand, otherwise we 
would need to make assumptions on the behavior of either the warp factor or the scalar field in order 
to get sensible results (see Sec. \ref{sec-fluct} for details). Notwithstanding, there is no guarantee 
of the fulfillment of these assumptions within a given scalar-tensor system. For instance, the 
assumption of certain behavior of the geometrical entities of the model could lead to a divergent 
behaviour of the scalar field and, hence, a divergent character of the effective 4D Planck mass, 
rendering an ill model where there is no localization of gravity. There is no need to mention that 
the complete and detailed analysis of the localization properties of {\it all} the tensor, vector 
and scalar perturbations of the scalar-tensor system that generates the braneworld model is very 
involved. Thus, another aim of this article is the rigorous study of the localization properties 
of {\it all}  the sectors of gravity and matter field fluctuations for the considered scenario 
with exact field configurations, generalizing previously obtained results in which only the 
tensor sector was taken into account (see \cite{pasipoularides, CQG29}).

In other words, as an application of the obtained exact braneworld field 
configurations we shall perform a study of the conditions under which we recover:
i) a well defined effective 4D Planck mass;
ii) a localized tensor zero mode fluctuation (graviton) which accounts for the observed 4D general relativity theory;
iii) the presence of vector and scalar fluctuations that do not \emph{considerably} alter the 
low energy phenomenology observed in 4D.
This is because general relativity predicts that in empty 4D Minkowski space--time the  only  relevant 
fluctuations come from the tensor sector (from the massless mode).
Thus, although in our scenario the scalar and vector  sectors  are non--trivial,  
their effects must be suppressed for a 4D observer on the brane\footnote{This reasoning is 
confirmed  by the fact that in a 5D thick braneworld model generated by a scalar field 
minimally coupled to gravity, the corrections to the Newton's law coming from the scalar modes 
are more suppressed than the corrections arising from the tensor modes \cite{nlscalar}.
}. 
The delocalized character of non--tensor modes provides a mechanism that protects  the 
low energy physics of the flat brane from  these unwanted effects. In order to achieve these aims, we shall consider just 
positive values for the non-minimal coupling function, a condition 
which guarantees an effective positive 4D Planck mass.

The paper is organized as follows: in Section \ref{setup} we present the action of the braneworld 
configuration along with the field equations for the proposed metric ansatz. In 
Section \ref{sec-planck-masses} we obtain the expression for the effective 4D Planck 
mass under dimensional reduction starting from the original 5D action. 
We then consider, in Section IV, a positive valued non--minimal coupling 
function for the scalar field by imposing the condition $L>0$, which 
guarantees a positive effective 4D Planck mass. In particular, we 
consider a concrete function which offers a clear dependence of the 
effective 4D Planck mass on the non-minimal coupling parameter, a fact 
that in turn allows us to easily compare to this magnitude in the minimally 
coupled case, when this parameter vanishes. Moreover, this form of the 
function $L$ also makes evident that the scalar field must be bounded, restricting 
the universe of mathematically available solutions to those which are physically 
meaningful field configurations from the 4D point of view. We also choose a suitable warp 
factor for the metric within this Section. We further impose the conditions for the 
solutions of the scalar field to possess a 
mirror symmetry along the extra dimension in Section V. 
In this way we study some general properties that we wish to 
be present in our model, essentially, to have a positive and finite 4D coupling 
constant (the Planck mass), to consider regular warp factors that avoid the presence 
of curvature singularities, and to select physically simple solutions among all the 
obtained scalar field configurations. In Section \ref{exact-sol} we
construct exact solutions for the highly non--linear differential equations for the scalar field, a
necessary step for providing a rigorous study of the consistency and localization properties of
the metric fluctuations of our braneworld configuration. Moreover, these solutions are essential
for analyzing the positiveness of the 4D Planck mass (an indispensable property of any viable
theory). In Section \ref{sec-fluct}  we further analyze the perturbations of the geometry which can be classified into
scalar, vector and tensor sectors according to the 4D Poincar\'e symmetry group. We establish that
both the scalar and vector sectors are not localized on the brane in contrast to the 4D
tensor massless mode. We finally summarize our
results and conclusions at the end of the paper.

\section{The model}\label{setup}

Let us explore a thick braneworld described by the following 5D action (a similar set up is studied in references \cite{andrianov, mg00}):
\bea S=-\int
d^5x\sqrt{|g|}\left[\frac{L}{2\kappa}R-\frac{1}{2}(\nabla\varphi)^2+V+\alpha' {\cal R}^2_{\rm
GB}\right],\label{action}\eea
where the constant $\alpha'>0$ and $\kappa \simeq 1/M^3$, $M$ -- the 5D
Planck mass. Besides, $\varphi$ is a real scalar field and $V=V(\varphi)$ its
self--interaction potential. The quantity $L=L(\varphi)$ describes the  non--minimal coupling
between the scalar field $\varphi$ and the Einstein--Hilbert term and  ${\cal R}^2_{\rm GB}$ is the
5D Gauss-Bonnet term (\ref{GB-term}).

In the scalar-tensor gravity theory (\ref{action}), 15 degrees of freedom $g_{AB}$, plus the
scalar field $\vphi$, propagate the gravitational interaction. Hence, this is not a pure
geometrical theory of gravity. In particular, the metric coefficients define the geodesic motion
of test particles, while the scalar field determines locally the strength of gravity by means of
the effective gravitational coupling $\propto\kappa/L(\vphi)$.

The way the scalar field $\vphi$ is coupled to the curvature in the present theory, deserves an
independent comment. Actually, in this paper, just as a matter of necessary simplicity, we have
considered explicit coupling of $\vphi$ to the curvature scalar $R$, but not to the Gauss-Bonnet
invariant. Intuition, instead, dictates that the scalar field should couple in a similar fashion
to $R$ and to ${\cal R}^2_{\rm GB}$, since both contribute towards the curvature of space-time.
This will be the subject of forthcoming work. We want to underline that this `asymmetric' coupling
has nothing to do with physical requirements, but just with simplicity of further mathematical
handling since, as we shall see further, the relevant equations of motion are highly non--linear.

The Einstein's field equations that come from the action (\ref{action}) take the
following form:
\be L\,R_{A B}+\epsilon\,{\cal Q} _{A B}= \kappa\,\tau_{A B}\,+\nabla_A\nabla_B
L+\frac{1}{3}\,g_{A B}\,\Box L,\label{ee}\ee
where
\bea
{\cal Q}_{AB}&=&\frac{1}{3}\,g_{A
B}\,{\cal R}^2_{\rm
GB}-2\,R\,R_{AB}+4\,R_{AC}\,R^{C}_{\,\;B} + 4R^{CD}\,R_{ACBD}-2\,R_{ACDE}\,R_{B}^{\,\;CDE},
\eea
is the Lanczos tensor representing the Gauss-Bonnet corrections to the Einstein's
field equations, $\epsilon=2\alpha'\kappa$, $\Box\equiv g^{CD}\nabla_C \nabla_D$, and the
stress-energy tensor $\tau_{A B}$, corresponding to the scalar matter content on the bulk, is
defined as usually:
$$\tau_{A B}=\partial_{A}\varphi\partial_{B}\varphi-\frac{2}{3}\,g_{A B}\,V(\varphi).$$ The
remaining terms in the right-hand-side (RHS) of (\ref{ee}) come from the non--minimal coupling of
the scalar field to the curvature scalar.

Furthermore, the equation that describes the dynamics of the scalar
field (Klein-Gordon equation) can be written as follows:
\be
\Box\varphi+\frac{1}{2\kappa}\,R\,L_{\varphi}+\frac{dV}{d\varphi}=0,\label{kg}\ee
where $L_{\varphi}=dL/d \varphi$. As it is done, for instance, in Ref. \cite{mg1}, the geometry of our braneworld model is described by a warped metric in conformally flat coordinates:
  \be ds^2=a^2(w)
[\eta_{\mu\nu}dx^{\mu}dx^{\nu}-dw^2],\label{line-e}
\end{equation}
where we use the signature $(+----)$, $\eta_{\mu \nu}$ is the 4D Minkowski metric, the variable $w$ is the extra coordinate and all the dimensions are of infinite extend.
Due to the Einstein field equations, the simplicity of our metric ansatz (\ref{line-e}) implies that the field $\varphi$ 
depends only on the fifth coordinate $w$. Thus, the 
Einstein and Klein--Gordon equations (\ref{ee}) and (\ref{kg}) give rise to
\bea
&&V+\frac{3\,{\cal H}\,L_{\varphi}\,\varphi'}{\kappa\,a^2}+
\frac{1}{2\kappa \,a^2}\biggl(\varphi''\,L_{\varphi}+\varphi'^2\,L_{\varphi\varphi}\biggr) 
-\frac{3}{2\kappa a^2}\biggl[\frac{4\,\epsilon}{a^2}\,{\cal H}^2\,\left({\cal H}^2+
{\cal H}'\right)-\left({\cal H}'+3{\cal H}^2\right)\,L\biggr]=0,\label{eequation1}\\
&&\frac{1}{\kappa}L_{\varphi}\,\varphi''-\frac{2}{\kappa}\,{\cal H}\,L_{\varphi}\,\varphi'+
\left(1+\frac{L_{\varphi\varphi}}{\kappa}\right)\varphi'^2-\frac{3}{\kappa}\left({\cal H}^2-
{\cal H}'\right)q=0,\label{eequation2}\\ 
&&\varphi''+3{\cal H}\varphi'-\frac{dV}{d\varphi} a^2-\frac{2L_{\varphi}}{\kappa}
\left( 3{\cal H}^2+2{\cal H}'\right)=0.\label{eequation3}
\eea
Here a prime denotes derivative with respect to 
the extra-coordinate $w$ (for instance, $\varphi'=d\varphi/dw$), in addition to this
$${\cal H}\equiv\frac{a'}{a},\qquad\ \mbox{while}\qquad
q\equiv L-\frac{4\,\epsilon}{a^2}\,{\cal H}^2.$$

As one can straightforwardly check, the three equations (\ref{eequation1})-(\ref{eequation3}) are 
not independent, only two of them are.

\section{Planck masses}\label{sec-planck-masses}

In this Section we shall derive the 4D effective coupling constant of our model, the Planck mass, starting from the 5D action (\ref{action}) through a mechanism called dimensional reduction. In general, the 4D effective theory can be obtained by integrating the 5D action with respect to the fifth coordinate $w$.
In order to do that let us consider a generalization of the 5D line--element (\ref{line-e}) where the 4D Minkowski metric $\eta_{\mu\nu}$
is replaced by an arbitrary 4D metric  $\tilde{g}_{\mu \nu}(x)$\footnote{On a general ground, the dimensional reduction of the original 5D action must render a non--trivial 4D Einstein--Hilbert term. Therefore, such a generalization of the 4D metric is needed because the curvature scalar corresponding to 4D Minkowski space--time vanishes.
 The simplest way to do that it is to replace $\eta_{\mu \nu} \mapsto \eta_{\mu \nu}+\delta g_{\mu \nu}(x)$, where $\delta g_{\mu \nu}(x)$ is a perturbation of the 4D Minkowski geometry.}:
\be ds^2 =a^2(w)[\tilde{g}_{\mu\nu}(x)dx^{\mu}dx^{\nu}-dw^2], \quad \mbox{where} \quad x\equiv \{x^{\mu}\}. \label{metric}
\ee
When performing the integration with respect to the extra dimension, the 5D fundamental theory
is reduced to a 4D Einstein-Hilbert effective action plus the corrections that come from the
scalar matter and higher curvature terms of the bulk.
In this Section we will focus in the analysis of
the 4D Planck mass, therefore  we  only need to
extract the 4D Einstein-Hilbert effective action after the  dimensional reduction of
the original 5D action
\begin{equation}
\label{EH_eff_action}
S_{4}\simeq M_{\rm Pl}^2\int d^{4} x
\sqrt{|\tilde{g}_4|}\tilde{R}_{4}+\cdots,
\end{equation}
where the subscript $4$ labels quantities computed
with respect to 4D metric $\tilde{g}_{\mu \nu}(x)$ and $M_{\rm Pl}$ is the effective 4D Planck mass.

A method for explicitly finding the 4D effective theory, in particular the 4D Einstein-Hilbert action, starting from the 5D one consists in interpreting the warp factor as a conformal function
as follows \cite{faraoni}
\be
\tilde{g}_{A B}\mapsto g_{A B}=a^2(w)\tilde{g}_{ A B}(x)=a^2(w) \left(\begin{array}{cc}
\tilde{g}_{\mu\nu}(x) & 0 \\ 0 & -1 \\ \end{array}\right),
\label{1}
\ee
and rewriting the 5D action  in terms of the quantities defined with respect to $\tilde{g}_{A B}(x)$. Eq.(\ref{EH_eff_action}) asserts the splitting of the quantity $\tilde{g}_{A B}(x)$ into $\tilde{g}_{\mu\nu}(x)$ and $g_{ww}=-1$.

In order to obtain the Einstein-Hilbert part of the 4D effective action, it is only necessary to consider the above conformal transformation on the terms
$-L(\varphi)\,R/2\kappa$ and $\alpha'{\cal R}^2_{\rm GB}$ in (\ref{action}).
The following expressions display these quantities after the conformal transformation
\bea
R&=&a^{-2}\left(\tilde{R}-8\tilde{\Box}\vartheta-12(\tilde{\nabla}\vartheta)^{2}\right),\label{ricci_jf}\\
{\cal R}_{GB}^2&=&\frac{1}{3 a^{4}}\left\{\tilde{\cal R}_{GB}^2-12\tilde{R} (\tilde{\nabla}
\vartheta)^2-24\tilde{R}\tilde{\Box}\vartheta +
48\tilde{R}^{AB}\left(\tilde{\nabla}_{A}\tilde{\nabla}_{B}\vartheta-
\tilde{\nabla}_{A}\vartheta\tilde{\nabla}_{B}\vartheta\right)\right.\nonumber\\
&+&72\left[(\tilde{\Box}\vartheta)^2+(\tilde{\nabla}\vartheta)^4-
(\tilde{\nabla}_{A}\tilde{\nabla}_{B}\vartheta)(\tilde{\nabla}^{A}\tilde{\nabla}^{B}\vartheta)\right]\nonumber\\
&+&\left. 144
(\tilde{\nabla}_{A}\tilde{\nabla}_{B}\vartheta)(\tilde{\nabla}^{A}\vartheta)(\tilde{\nabla}^{B}
\vartheta) +144 (\tilde{\nabla} \vartheta)^2 \tilde{\Box} \vartheta\right\},\label{gb_jf}
\eea
where
$\vartheta=\ln a$. The terms  with a tilde are defined with respect to the metric $\tilde{g}_{A
B}(x)$. The next step consists in using the relations (\ref{1})--(\ref{gb_jf}) in order to obtain the effective 4D Ricci scalar. Therefore, after substituting (\ref{ricci_jf}) and (\ref{gb_jf}) into (\ref{action}) and integrating the prefactor of the 4D Ricci scalar with respect to the fifth dimension, the relation between 4D and 5D Planck masses can be written as follows
\bea
&&M_{\rm Pl}^2\simeq M^3\int_{-\infty}^{\infty} a^3(w)\left[L+\frac{4\epsilon}{a^2}
({\cal H}^2 + 2{\cal H}')\right] dw\nonumber\\
&&\;\;\;\;\;\;\;=M^3\int_{-\infty}^{\infty}
a^3(w)\,q\,dw+8\,M^3\,\epsilon\,[a']_{-\infty}^{\infty}.
\label{masaplanck}
\eea
It is evident that the $M_{\rm Pl}$ explicitly depends on the non--minimal coupling function $L$
as well as on the Gauss--Bonnet corrections to the Einstein--Hilbert term parameterized by $\epsilon$.

This quantity should be positive and finite for a consistent theory. Moreover, it should reproduce
the effective gravitational couplings that we observe in our 4D world, if we wish to recover 4D
gravity on the brane\footnote{In particular, if the non-minimal coupling funtion $L(\varphi)$ increases with 
the same rapidity as (or faster than) the factor $a^{-3}(w)$, then the effective 4D Planck 
mass $M_{Pl}$ will diverge, leading to 
a theory with unphysical 4D gravitational couplings.}. Later on we shall see that these 
conditions are fulfilled for a wide class
of solutions of our scalar--tensor model.

\section{Background Geometry and Non--minimal Coupling}\label{wf_nmc}

Although our scenario does not impose any
restriction on the functional form of $L(\varphi)$, apart of being positive,
here we shall consider the simple non trivial  coupling function \cite{farakos} (see also \cite{andrianov}):
\be 
L(\varphi)=1-\frac{\xi}{2}\,\varphi^2.\label{l}
\ee
The choice of the coupling function $L(\varphi)$ in the above expression is 
inspired in 4D phenomenology \cite{brans}--\cite{lee}.
The parameter $\xi$
characterizes the strength
of the  non--minimal
coupling between scalar matter and gravity.
Besides, if assume small values of the coupling parameter  -- as it is implicit in the present 
paper -- and bounded scalar, this choice 
reflects the fact that only small deviation from the case $\xi=0$ is being considered.
When  $\xi=0$, the scenario is described by a bulk scalar field minimally coupled to gravity
(it was explored in Ref. \cite{mg00}).
This particular form of $L(\varphi)$ is very
simple, however, it still gives us an idea of the effects
of the non--minimal coupling on the thick brane, furthermore,
in contrast with other more complicated ansatze, it is not too
hard to solve its associated field equations. Notwithstanding, it is worth  noticing that one can consider 
different non--minimal coupling functions $L(\vphi)$ which still are positive and finite that do not 
restrict the scalar field and lead to interesting results like deformed braneworld configurations as 
in \cite{liuNM}. The important point here is to avoid getting negative or divergent values for the integral
$M_{Pl}^2\sim \int L(\vphi(w))a^3(w)dw$ since this would imply an effective 4D coupling constant that 
does not reproduce the observed gravitational interactions of our world.

Positivity of $L$ leads to $\vphi^2$ being bounded from above:
\be L>0\quad\;\Rightarrow\quad\;\vphi^2<\frac{2}{\xi}\quad\;\Rightarrow\quad\;|\vphi|<\sqrt\frac{2}{\xi}
\label{bound}
\ee
and we will impose this restriction by hand on the field configurations obtained below in order to ensure a positive and finite value for the effective 4D Planck mass.

In this work, besides, we shall consider a regular geometry of the form (\ref{line-e}), which interpolates
between two asymptotically $AdS_5$ space-times, depicted by the following warp factor
\cite{mg00}:
\be a(w)=\frac{a_0}{\sqrt{1+b^2 w^2}},\label{warp-f}
\ee
where the  width of the
thick brane is  $1/b$, and  the quantity $a_0$ is dimensionless and related to the radius of the asymptotic $AdS_5$ space-time. All
of the resulting quadratic curvature invariants constructed with this warp factor are regular and asymptotically constant.

Once we have introduced the non--minimally coupling function (\ref{l}) and the warp factor (\ref{warp-f}), we can proceed to solve 
the field equations (\ref{eequation1})-(\ref{eequation3}). From the mathematical point of view, these differential equations are conveniently treated in terms of a dimensionless variable. 
Therefore, we shall perform the following change of variable $w \mapsto v=bw$, where $v$ is
dimensionless. Thus, by using (\ref{l}) and (\ref{warp-f}), Eq. (\ref{eequation2}) can be rewritten as:
\begin{eqnarray}
&&\xi\varphi\varphi''+\frac{2\xi \,v}{1+v^2}\,\varphi\varphi'
+(\xi-\kappa)\varphi'^{2}-\frac{3\xi}{2}\frac{\varphi^2}{(1+v^2)^2}
=-\frac{3}{(1+v^2)^2}+\frac{4\epsilon b^2}{a_0^2}\frac{3v^2}{(1+v^2)^3},\label{master-eq}
\end{eqnarray}
where, now, the prime denotes derivative with respect to the dimensionless variable $v$.
It is difficult to find the general solution to the above equation, however, several interesting (particular)
exact solutions can be found. Here we shall split the analysis into three separated cases
corresponding to: i) a scalar field minimally coupled to gravity and including a 5D Gauss-Bonnet
invariant ($\xi=0$ and $\epsilon\neq 0$) \cite{mg00}, ii) a scalar field non--minimally coupled to
curvature without the Gauss-Bonnet term ($\xi\neq 0$ and $\epsilon=0$), and iii) the more general
situation when both non--linear effects are present, i.e. when $\xi\neq 0$ and $\epsilon\neq 0$.
Once a given particular (exact) solution for the scalar field $\vphi=\vphi(v)$ is found,  one
hence can write the self-interaction potential as a function of the (dimensionless) extra-dimensional
coordinate $v$:
\bea
V(v)&=&\frac{3b^2}{2\kappa a_0^2}\left\{\frac{4\epsilon
b^2}{a_0}\frac{v^2(2v^2-1)} {(1+v^2)^2}+\frac{1-4v^2}{1+v^2}+
\xi\left[(1+v^2)(\vphi\vphi''+\vphi'^2)-2v\vphi\vphi'-\left(\frac{1-4v^2}{1+v^2}\right)\frac{\vphi^2}{2}\right]\right\}.\label{pot}
\eea
The above expression is just a rewriting of Eq.  (\ref{eequation1}) for the case of interest
in this paper, i.e. for the non--minimally coupling function (\ref{l}) and the warp factor (\ref{warp-f}).

It is worth mentioning that only when we have $\vphi(v)$ and $a(v)$ at hand, we can perform the integration of (\ref{masaplanck}) and compute the effective 4D Planck mass $M_{\rm Pl}^2$. Moreover, once $\vphi(v)$ and $V(\vphi(v))$ are given as functions of the dimensionless variable $v$, the components of the stress energy tensor $\tau_{AB}=\tau_{AB}(v)$, are also known functions of
the dimensionless extra-dimensional variable.

\section{Mirror Symmetry}\label{m-sym}

An important remark on the particular solutions we are looking for is related to the symmetries
of the metric coefficients. As mentioned, in the general case when there is a non--minimal coupling
between the scalar field and the curvature scalar ($\xi\neq 0\;\Rightarrow$ $L=L(\vphi)\neq 1$),
gravity is propagated both by $g_{AB}$ and by $\vphi$. Hence, one should naively expect that the
(real) scalar field, $\vphi(v)$, will respect the same symmetries as the metric functions
$g_{AB}$. In the remaining part of this Section we shall show that, as a matter of fact, while the
latter is a mandatory requirement for the case when $\xi=0$ (in the minimal coupling case), this
property is not present in the general case in which $\xi\neq 0$.

In the present case the metric coefficients in (\ref{line-e}) inherit the `mirror' symmetry
\be
v\mapsto -v\quad\;\Rightarrow\quad\;g_{AB}(v)=g_{AB}(-v),\label{mirror}
\ee
distinctive of the warp factor (\ref{warp-f}). In consequence, the tensors $R_{AB}$, and ${\cal Q}_{AB}$, in the
left-hand-side (LHS) of Einstein's field equations (\ref{ee})
\be
R_{AB}(v)=R_{AB}(-v),\qquad\quad\;{\cal Q}_{AB}(v)={\cal Q}_{AB}(-v),\label{mirror'}
\ee
also respect invariance under (\ref{mirror}).

For the minimal coupling case ($\xi=0$ $\Rightarrow L=1$), since the Einstein's field equations read,
\be R_{AB}+\epsilon{\cal Q}_{AB}=\kappa\,\tau_{AB},\label{ee-minimal}\ee the invariance
(\ref{mirror'}) will entail that, necessarily, $\tau_{AB}(v)=\tau_{AB}(-v)$, which means, in turn,
that
\be V(v)=V(-v),\quad\qquad\;\vphi(v)=\pm\vphi(-v),
\label{mirror-vphi}
\ee
where, in the last equality, only one of the two signs, ``$+$'', or ``$-$'', is to be chosen at once. In other words,
$\vphi$ can be either even or odd, under (\ref{mirror}).

In the general case $\xi\neq 0$ ($L=L(\vphi)\neq 1$), on the contrary, since the coupling function
$L$ is multiplying the Ricci tensor in the LHS of (\ref{ee}), mirror symmetry (\ref{mirror}),
(\ref{mirror'}) is respected only if the additional requirement
\be
L(\vphi(v))=L(\vphi(-v)),\label{mirror''}
\ee
is fulfilled.\footnote{In our analysis we took into
account the fact that the operators $\nabla_A\nabla_B$, and $\Box$ in the RHS of Eq. (\ref{ee}),
are invariant under (\ref{mirror}). Recall that, in the present case, there is functional
dependence on the extra-coordinate $v$ only.} In the case studied in this paper, since $L$ is
quadratic in $\vphi$ (see Eq. (\ref{l})), the latter requirement will entail, again, that, under
(\ref{mirror}), $\vphi$ can be either even, or odd.

In case $\vphi$ does not show any obvious symmetry under (\ref{mirror}),
$$\vphi(v)\neq\pm\vphi(-v)\;\qquad\Rightarrow\qquad\;L(\vphi(v))\neq L(\vphi(-v)),$$ the coupling function $L$
counteracts the symmetry displayed by $R_{AB}$ and, in consequence, the LHS of (\ref{ee}) -- the
pure geometrical part of Einstein's equations -- is not invariant under (\ref{mirror}) anymore.
Hence, if $\xi\neq 0$, mathematical consistency of the solutions does not impose any mirror
symmetry requirement on $\vphi$. This can be only an independent `ad hoc' requirement.

Notwithstanding, in the present paper the symmetry requirements (\ref{mirror}), (\ref{mirror'}),
(\ref{mirror''}) will be used to select physically simple solutions among the set of all
possible exact solutions we will be able to find. Consequently, we shall take into consideration
only those $\vphi$-configurations which are either even ($\vphi(v)=\vphi(-v)$) or odd
($\vphi(v)=-\vphi(-v)$) under $v\mapsto -v$. While for the minimally coupled case it is
legitimate, as shown above, in the case when $\xi\neq 0$, it is not required by mathematical
consistency, but, by symmetric aesthetic, instead.

\section{Exact solutions}\label{exact-sol}

In this Section we shall exactly solve the differential Eq. (\ref{master-eq}) for certain special cases and present graphics of the corresponding profiles of the self--interaction potential given by (\ref{pot}).

As already mentioned, here we shall split the study into three particular cases:

A) $\epsilon\neq 0$ and $\xi=0$ -- minimal coupling.

B) $\xi\neq 0$ and $\epsilon=0$ -- non-minimal coupling.

C) $\epsilon\neq 0$ and $\xi\neq 0$ -- the general case.

In each case we will rely on a number of mathematical assumptions so that, after achieving considerable simplification, particular exact solutions can be found.

\subsection{Minimal Coupling Case ($\epsilon\neq 0$ and $\xi=0$)} \label{case-i}

If one sets $\xi=0$ then (\ref{master-eq}) adopt the following form
\be
\varphi'^{2}=\frac{3}{\kappa(1+v^2)^2}-\frac{4\epsilon b^2}{\kappa a_0^2}\frac{3v^2}{(1+v^2)^3}
\label{master-eqxi=0}
\ee
and can be easily solved. Actually, let us first change to the variable
\be
\tau=\arctan(v)\quad\;\Leftrightarrow\quad\;v=\tan(\tau),\quad \mbox{hence} \quad
\;\sin^2\tau=\frac{v^2}{1+v^2},
\label{tau-x}\ee
so that the derivatives of $\varphi$ with respect to $v$ and $\tau$ are related by
\bea
(1+v^2)\frac{d}{dv}=\frac{d}{d\tau}\quad\;\Leftrightarrow\quad\;(1+v^2)\varphi'=\dot\varphi,
\label{dertau-x}\eea
where overdots stand for derivatives with respect to $\tau$.
Notice that $v\in\;]-\infty,+\infty[\;\Rightarrow\,\tau\in\;]-\pi/2,\pi/2[$. In terms of this new
variable, Eq. (\ref{master-eqxi=0}) transforms into the following first order
equation: \be \dot\vphi=\pm\sqrt\frac{3}{\kappa}\sqrt{1-\frac{4\epsilon
b^2}{a_0^2}\sin^2\tau},\label{master-eq-i}\ee which can be integrated in quadratures to give:
\be
\vphi^\pm(\tau)=\pm\sqrt\frac{3}{\kappa}\,E(\sin\tau,k) +\vphi^\pm_0,\label{sol-i-1}
\ee
where
\be
E(\sin\tau,k)=\int_0^\tau\sqrt{1-k^2\sin^2\zeta}\,d\zeta,\qquad\qquad\;k=\sqrt\frac{4\epsilon b^2}{a_0^2},
\ee
is the incomplete elliptic integral of the second kind \cite{elliptic-i}, $\vphi^\pm_0$ are
arbitrary integration constants, and the $\pm$ signs represent different branches of the solution.

\begin{figure}[htb]
\begin{center}
\includegraphics[width=8cm]{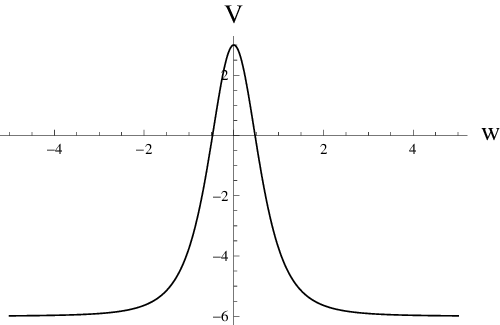}
\end{center}
\caption{The self--interaction potential for the case $\epsilon\neq 0$ and $\xi=0$. In this figure we set $k=1$, $a_0=1$, $b=1$ and  $\kappa=1/2$.}
\label{v_giovannini}\end{figure}

The particular case when $a_0=\sqrt{4\epsilon}b\;\Rightarrow\;k=1$ was studied in \cite{mg1}, where
$\vphi^\pm(\tau)=\pm\sqrt{3/\kappa}\;\sin\tau+\vphi^\pm_0$, or, in terms of the dimensionless
variable $v$ (see Eq. (\ref{tau-x})):
\be
\varphi^\pm(v)=\pm\sqrt\frac{3}{\kappa}\frac{v}{\sqrt{1+v^{2}}}+\varphi^\pm_0.\label{sol-i-1'}\ee
As before, the $\pm$ signs describe two possible branches of the solution.
The scalar field solutions (\ref{sol-i-1}),
(\ref{sol-i-1'}), do not respect invariance under (\ref{mirror-vphi}) unless the constants
$\vphi^\pm_0$ are set to zero: $\vphi^\pm_0=0$\footnote{Recall that
$E(\sin\tau,k)=-E(-\sin\tau,k)$.}.
Hence, in order to meet the wished symmetry requirements mentioned in the Section \ref{m-sym},
the particular exact solutions of Eq. (\ref{master-eq}), for the minimally
interacting case ($\xi=0$), are:
\bea &&\vphi^\pm(v)=\pm\sqrt\frac{3}{\kappa}
\,E\left(\frac{v}{\sqrt{1+v^2}},k\right),\;\;\;k\neq 1,\label{kne1}\\
&&\vphi^\pm(v)=\pm\sqrt\frac{3}{\kappa}\,\frac{v}{\sqrt{1+v^2}},
\;\;\;\;\;\;\;\;\;\;\;\;\;\;\;\;k=1.\label{sol-i}
\eea
The profile of the self--interaction potential $V$ is shown in Fig. (\ref{v_giovannini}).
As one can see, $V$ is asymptotically
constant and negative.
 This result is consistent with the fact that our geometry (\ref{line-e}) is asymptotically $AdS_5$.

Thus, the 4D Planck mass corresponding to the minimally coupled case with Gauss-Bonnet term is
\be
M_{Pl}^2 = \frac{a_{0}^3}{2b\kappa}-\frac{4 a_0 b\epsilon}{3\kappa}=\frac{a_{0}^3}{2b\kappa}\left(1-\frac{2k^2}{3}\right).
\label{Mpl6.1}
\ee
This is a finite quantity where the correction coming from the Gauss-Bonnet invariant is encoded in the second term of the RHS.

\subsection{Non-minimal Coupling Case ($\xi\neq 0$ and $\epsilon=0$)}\label{case-ii}

Now when $\epsilon=0$ in (\ref{master-eq}) we obtain the following differential equation
\begin{eqnarray}
\varphi\varphi''+\frac{2\,v}{1+v^2}\,\varphi\varphi'
+\left(1-\frac{\kappa}{\xi}\right)\varphi'^{2}-\frac{3}{2}\frac{\varphi^2}{(1+v^2)^2}=-\frac{3}{\xi(1+v^2)^2}.
\label{master-eqe=0}
\end{eqnarray}

If we make the same replacement as before $v\mapsto\tau=\arctan v$ (see Eq. (\ref{tau-x})) when $\epsilon=0$, then Eq. (\ref{master-eqe=0}) can be rewritten in the form
\be
\vphi\ddot\vphi+\left(\frac{\xi-\kappa}{\xi}\right)\dot\vphi^2
=\frac{3}{2}\vphi^2-\frac{3}{\xi}.
\label{master-eq-ii}\ee Then, without loss of generality, one
can make the following assumption: \be \dot\vphi^2=h(\vphi)\quad\Rightarrow\quad\ddot\vphi=
\frac{1}{2}\frac{dh(\vphi)}{d\vphi},\label{ansatz}\ee so that Eq. (\ref{master-eq-ii}) can be
written as a first-order differential equation,
\be
\frac{dh(\vphi)}{d\vphi}+2\lambda\frac{h(\vphi)}{\vphi}=
3\vphi-\frac{6}{\xi\vphi},\label{1-order-eq}\ee where
\be
\lambda=\frac{\xi-\kappa}{\xi}.
\label{lambda}
\ee
The following expression for $h(\vphi)$ solves Eq. (\ref{1-order-eq}): \be
h(\vphi)=\frac{3\vphi^2}{2(1+\lambda)}- \frac{3}{\xi\lambda}+C\,\vphi^{-2\lambda},
\label{sol-1-order-eq}\ee where $C$ is an arbitrary constant.


Here, for the sake of simplicity of mathematical handling, we set $C=0$ in Eq.
(\ref{sol-1-order-eq}). Then, after substituting back (\ref{sol-1-order-eq}) into (\ref{ansatz}),
one is left with the following first order differential equation:
\be
\dot\vphi=\pm\sqrt\frac{3}{2(1+\lambda)}
\sqrt{\vphi^2-\frac{2(1+\lambda)}{\xi\lambda}}.
\label{1-master-eq-ii}
\ee
Eq. (\ref{1-master-eq-ii}) can be integrated in quadratures:
\be
\pm\int\frac{d\vphi}{\sqrt{\vphi^2-\frac{2}{\xi}\left(\frac{2\xi-\kappa}{\xi-\kappa}\right)}}=
\sqrt\frac{3\xi}{2(2\xi-\kappa)}\;\tau+C_0, \label{quad-ii}
\ee
where we have returned to the
original parameters $\xi$ and $\kappa$ through (\ref{lambda}), and $C_0$ is an integration constant which in the following
calculations we fix to meet the imposed symmetry requirement (\ref{mirror-vphi}). Hence, depending on the interval in $\xi$-parameter, one will obtain different particular exact solutions:

\begin{enumerate}

\item Case $0<\xi<\kappa/2$
\be
\vphi^\pm(\tau) = \pm\, \sqrt\frac{2(\kappa-2\xi)}{\xi(\kappa-\xi)}\;
\cos\left(\sqrt\frac{3\xi}{2(\kappa-2\xi)}\;\tau\right),\label{sol-ii-1}
\ee
or, in terms of the original dimensionless variable (see Eq. (\ref{tau-x})):
\be
\vphi^\pm(v) = \pm\, \vphi_{01}\cos\left(\beta\,\arctan v\right),\label{sol-ii-1'}
\ee
where, for
compactness of writing, we have introduced the following constants:
\be
\vphi_{01}=\sqrt\frac{2(\kappa-2\xi)}{\xi(\kappa-\xi)},\quad\qquad
\;\beta=\sqrt\frac{3\xi}{2(\kappa-2\xi)}.
\ee
The bound (\ref{bound}) on $\vphi^2$ leads to:
$$
\vphi_{01}^2=\frac{2(\kappa-2\xi)}{\xi(\kappa-\xi)}<\frac{2}{\xi}\qquad\;\Rightarrow
\qquad\;
\kappa-2\xi<\kappa-\xi,
$$
which, for $0<\xi<\kappa/2$, is always fulfilled. Hence,
positivity of $L(\vphi)$ does not impose any additional constraint on the parameter $\xi$.

On the other hand, for the above profile of the field the self--interaction potential is constant 
and negative at $w \rightarrow \infty$ (see Fig. \ref{V_e=0_chi_nocero_1sol}).
\begin{figure}[htb]
\begin{center}
\includegraphics[width=8cm]{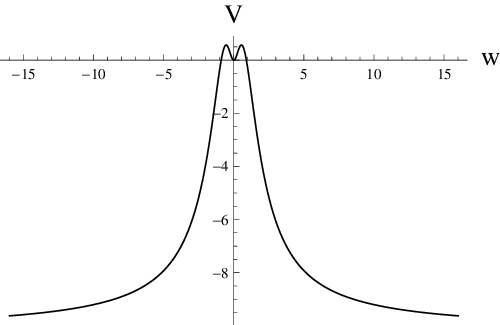}
\end{center}
\caption{$V$ corresponding to solution (\ref{sol-ii-1'}). In this
figure we set $\xi=1/10$, $a_0=1$, $b=1$ and  $\kappa=1/2$.}
\label{V_e=0_chi_nocero_1sol}\end{figure}

The corresponding effective 4D Planck mass for this case reads
\be
M_{Pl}^2 = \frac{a_{0}^3}{2b\left(\kappa-\xi\right)}\,\left[1+\frac{\left(\kappa-2\xi\right)^2\cos\left(\beta\pi\right)}{\kappa\left(8\xi-\kappa\right)}\right],
\label{Mpl6.2.1}
\ee
which is finite and positive since $L>0$ within the interval $0<\xi<\kappa/2$.

\item Case $\kappa/2<\xi<\kappa$
\be \vphi^\pm(\tau) = \pm\, \sqrt\frac{2(2\xi-\kappa)}{\xi(\kappa-\xi)}\;
\sinh\left(\sqrt\frac{3\xi}{2(2\xi-\kappa)}\;\tau\right).\label{sol-ii-2}
\ee
This solution can be expressed in the language of the dimensionless variable $v=b w$:
\be
\vphi^\pm(v) = \pm\, \vphi_{02}\sinh(\beta\,\arctan v),\label{sol-ii-2'}
\ee
where
$$\vphi_{02}=\sqrt\frac{2(2\xi-\kappa)}{\xi(\kappa-\xi)}.$$

The positive character of $L$ within the above mentioned interval, $\kappa/2<\xi<\kappa$, renders the following condition for any value of $\tau$:
\be
\sinh^2\left(\sqrt\frac{3\xi}{2(2\xi-\kappa)}\;\tau\right)<\left(\frac{\kappa-\xi}{2\xi-\kappa}\right).
\label{ineq2}
\ee
In principle, this inequality can be fulfilled around the point where $\tau$ vanishes
\be
0<\left(\frac{\kappa-\xi}{2\xi-\kappa}\right).
\ee
Notwithstanding, when $\tau$ approaches the values $\pm\frac{\pi}{2}$, the inequality (\ref{ineq2}) becomes
\be
\sinh^2\left(\sqrt\frac{3\xi}{2(2\xi-\kappa)}\;\frac{\pi}{2}\right)<\left(\frac{\kappa-\xi}{2\xi-\kappa}\right).
\ee
It turns out that this inequality can never be satisfied within the interval $\kappa/2<\xi<\kappa$. 
This conclusion can be made from the behavior of $L$ within the above mentioned 
interval: while the LHS of the inequality 
exponentially diverges as $\xi\mapsto\frac{\kappa^+}{2}$ and evolves towards the 
value $\sinh^2\left(\sqrt\frac{3}{2}\;\frac{\pi}{2}\right)$ 
as $\xi\mapsto\kappa^-$, the RHS linearly diverges as $\xi\mapsto\frac{\kappa^+}{2}$ and 
vanishes as $\xi$ approaches the other end of the 
interval, i.e. when $\xi\mapsto\kappa^-$. Thus, the non-minimal coupling function $L$ is not 
definite positive and possesses regions where it is negative within the interval $\kappa/2<\xi<\kappa$.\footnote{ It 
is worth noticing that one can consider that $L(\varphi)>0$ implies that the extra dimension 
is compact. In this case the Planck mass is finite. However, since we are considering an 
unbounded extra dimension, we shall not consider this situation here.}

Since $L>0$ is a very strong condition, in principle it could be possible to still have 
a positive Planck mass for a more restricted interval of $\xi$.
This fact indicates that one must compute the effective 4D Planck mass (relaxing 
for a while the $L>0$ condition) in order to see whether the correct gravitational 
couplings can still be recovered in 4D for the aforementioned solution 
of the model within the interval $\kappa/2<\xi<\kappa$:
\be
M_{Pl}^2 = \frac{a_{0}^3}{2b\kappa}	\left[\frac{\kappa}{\left(\kappa-\xi\right)}-\frac{\left(2\xi-\kappa\right)^2}{\left(\kappa-\xi\right)}
\frac{\cosh\left(\beta\pi\right)}{\left(8\xi-\kappa\right)}\right].
\label{Mpl6.2.2}
\ee
One must further impose the finiteness and positivity of the 
effective 4D Planck mass, requirements that lead to the following 
condition for $\kappa/2<\xi<\kappa$:
\be
\frac{\kappa\left(8\xi-\kappa\right)}{\left(2\xi-\kappa\right)^2}>
\cosh\left(\beta\pi\right).
\label{ineqcosh}
\ee
This inequality cannot be fulfilled when $\xi$ lies within 
the interval $\kappa/2<\xi<\kappa$: the RHS of the inequality diverges 
exponentially as $\xi\mapsto\frac{\kappa^+}{2}$ and approaches 
the value $\cosh\left(\sqrt\frac{3}{2}\;\pi\right)\approx 23.45$ when
$\xi\mapsto\kappa^-$, while the LHS diverges quadratically 
as $\xi\mapsto\frac{\kappa^+}{2}$ and tends to $7$ when
$\xi\mapsto\kappa^-$. Thus, the opposite claim actually 
holds: when $\xi$ lies between $\kappa/2$ and $\kappa$, then
$M_{Pl}^2<0$, yielding a negative 4D effective coupling constant 
for gravitational interactions, i.e., to repulsive 4D gravity.

\item Case $\xi>\kappa$

\be
\vphi^\pm(\tau) = \pm\, \sqrt\frac{2(2\xi-\kappa)}{\xi(\xi-\kappa)}\;
\cosh\left(\sqrt\frac{3\xi}{2(2\xi-\kappa)}\;\tau\right),\label{sol-ii-3}
\ee
or, in terms of $v=b w$,
\be
\vphi^\pm(v) = \pm\, \vphi_{02}\cosh(\beta\,\arctan v)\label{sol-ii-3'}.
\ee
However, positivity of the coupling function $L(\vphi)$ imposes the following non-algebraic
constraint on $\xi$ and $\kappa$:
\be
\cosh^2\left(\frac{\pi}{2}\sqrt\frac{3\xi}{2(2\xi-\kappa)}\right)<\left(\frac{\xi-\kappa}{2\xi-\kappa}\right).
\ee
This inequality is not compatible with the above assumed condition $\xi>\kappa$ since then, the LHS is always greater than one, while the RHS is less than the unity. It turns out that under the restriction $\xi>\kappa$, the function $L$ is negative along the whole extra dimension, giving rise a negative 4D effective Planck mass and, hence, to a repulsive gravity as in the previous case.

\end{enumerate}

All the non-minimally coupled scalar field solutions (\ref{sol-ii-1'}), (\ref{sol-ii-2'}) and (\ref{sol-ii-3'}) respect
the `mirror' symmetry (\ref{mirror-vphi})-(\ref{mirror''}), moreover, the expressions (\ref{sol-ii-1'}) and (\ref{sol-ii-3'})
are symmetric under $v\mapsto -v$: $\vphi^\pm(v)=\vphi^\pm(-v)$, whereas solution (\ref{sol-ii-2'}) is odd under such a symmetry: $\vphi^\pm(v)=\vphi^\mp(-v)$. However, from the above presented exact field configurations, just the solution (\ref{sol-ii-1'}) renders a 
physically viable model that can correctly describe the 4D effective gravity of our world.

\subsection{General Case: $\xi\neq 0$, $\epsilon\neq 0$}\label{case-iii}

In this case we shall construct solutions that involve both the non--minimal coupling of the scalar
field to gravity and the presence of the Gauss-Bonnet term. Despite the fact of the highly
non--linear character of the differential equation (\ref{master-eq}),
we were able to obtain some exact solutions when there exists some relationship between the parameters of our model.

\subsubsection{Particular solution: case $\lambda=-1$.}\label{casew=-1}

If one performs the following change of variable $y=\mbox{arcsinh}\,v$ and redefine the field $\varphi=\psi^{1/(1-\lambda)}$, then (\ref{master-eq}) can be easily
solved for the particular case when $\lambda=-1$  (see (\ref{lambda}) for reference) or, equivalently, $\kappa=2\xi$, since in this case we get a second order linear differential equation.

Let us see how this comes about: by changing the variable $v=\sinh
y$, hence, $\cosh y=\sqrt{1+v^2}$, from (\ref{master-eq}) we get the following equation for $\varphi$:
\be \!\!\!\!\!\!\!\!\!\!\!\! \varphi\left(\ddot\varphi+\tanh y\ \dot\varphi\right) - \lambda\left(\dot\varphi\right)^2 -
\frac{3}{2}\ \mbox{sech}^2y\ \varphi^2 = \frac{3}{\xi}\ \mbox{sech}^2y\ \left(\frac{4\epsilon
b^2}{a_0^2}\tanh^2y\!-\!1\right),
 \label{master-eq-y}
 \ee
where overdots now mean derivatives with respect to $y$.
By further performing the following substitution $\varphi=\psi^{1/(1-\lambda)}$ we get
\be \!\!\!\!\!\!\!\!\!\!\!\!\ddot\psi+\tanh y\ \dot\psi-\frac{3}{2}\left(\!1\!-\!\lambda\!\right)\mbox{sech}^2y\ \psi\!=\!\frac{3}{\xi}\left(\!1\!-\!\lambda\!\right)\mbox{sech}^2y\ \left(k^2\tanh^2y\!-\!1\right)\psi^{\frac{\lambda+1}{\lambda-1}},
\label{master-eq-psi}
\ee
where $k^2=\frac{4\epsilon b^2}{a_0^2}$. By choosing the particular case $\lambda=-1$ we actually get the
linear differential equation
\be \ddot\psi+\tanh y\ \dot\psi - 3\mbox{sech}^2y\ \psi = \frac{6}{\xi}\mbox{sech}^2y\
\left(k^2\tanh^2y-1\right). \label{eqnw=-1} \ee
This equation has the following real solution:
\be \psi = \varphi^2 =
\frac{14-10k^2+6k^2\mbox{sech}^2y}{7\xi}+C_1\cosh\left(A\right)+C_2\sinh\left(A\right),
\label{solnw=-1} \ee
where $A=2\sqrt{3}\arctan\left(\tanh\frac{y}{2}\right)$, and $C_1$ and $C_2$ are arbitrary
constants. By going back to the dimensionless variable $v$ we get the solution for the field $\varphi$:
\be
\varphi = \sqrt{\frac{14-10k^2}{7\xi}+\frac{6k^2}{7\xi\left(1+v^2\right)}+
C_1\cosh\left(A\right)+C_2\sinh\left(A\right)},
\label{solnw=-1inx}
\ee
where now $A=2\sqrt{3}\arctan\left(\frac{\sqrt{v^2+1}-1}{v}\right)$.

\begin{figure}[htb]
\begin{center}
\includegraphics[width=8cm]{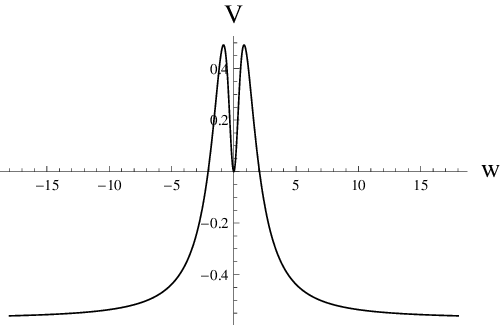}
\end{center}
\caption{$V$ associated to solution (\ref{solnw=-1}). We set $C_1=0$, $C_2=0$, $\xi=1/4$, $a_0=1$, $b=1$, $\epsilon=1/3$ and  $\kappa=1/2$.}
\label{V_generalsol1}\end{figure}

This solution fulfills the symmetry requirements (\ref{mirror-vphi})-(\ref{mirror''}) when $C_2=0$.
Similarly to previous cases, $V$ is asymptotically constant as it is shown in Fig. \ref{V_generalsol1}.

The positive nature of the radicand in the solution (\ref{solnw=-1inx}) restricts the values $C_1$ can adopt from below
\be
C_1 > \frac{2\left(2k^2-7\right)}{7\xi},
\label{C1below}
\ee
whereas by imposing the $L>0$ requirement, the integration constant $C_1$ is bounded from above
\be
C_1 < \frac{10k^2}{7\xi}\mbox{\rm sech}\left(\frac{\sqrt{3}\pi}{2}\right),
\label{C1above}
\ee
leading to the following constraint
\be
\frac{2\left(2k^2-7\right)}{7\xi}<C_1 < \frac{10k^2}{7\xi}\mbox{\rm sech}\left(\frac{\sqrt{3}\pi}{2}\right),
\label{C1}
\ee
which holds for  arbitrary $\xi$ and restricts the values of $k$ through $k^2<\frac{7}{2-5\mbox{\rm sech}\left(\frac{\sqrt{3}}{2}\pi\right)}$. Note that, in principle, $C_1$ can be negative.

When computing the 4D effective Planck mass according to (\ref{masaplanck}), we could not perform the integration for $C_1\neq 0$ and, hence, we have set this constant to zero. For this special case the Planck mass adopts the form
\be
M_{Pl}^2 = \frac{a_{0}^3}{21b\kappa}\left(9a_0^2k^2-28\epsilon b^2\right)=\frac{2a_{0}^3k^2}{21b\kappa},
\label{Mpl3.1}
\ee
which is positive definite and finite as it should be for a well defined 4D effective theory that reproduces the gravitational interactions of our world.

\subsubsection{Solutions for arbitrary $\lambda$.}\label{case_arb_w}

Let us recall that if we perform the coordinate transformation $x\mapsto\tau=\arctan v$ (see
Eq. (\ref{tau-x})) for the case in which the Gauss-Bonnet term is non--trivial, then the
Eq. (\ref{master-eq}) adopts the form:
\be \vphi\ddot\vphi+\left(\frac{\xi-\kappa}{\xi}\right)\dot\vphi^2
=\frac{3}{2}\vphi^2-\frac{3}{\xi}\left(1-k^2\sin^2\tau\right),\label{master-eq-iii} \ee
which will be from now on the relevant differential equation to be solved.

Let us now consider more general solutions in which the value of the parameter $\lambda$ is
arbitrary (see (\ref{lambda})). When dealing with Eq. (\ref{master-eq-iii}), we can perform the following
transformation in order to reduce the order of this differential equation
\be \dot\varphi(\tau)=\varphi(\tau)f(\tau), \label{phidot} \ee
where $f(\tau)$ is an integrable function of $\tau$. We further can redefine the scalar field
$\varphi$ as follows
\be
\varphi^2=\psi(\tau). \label{phicuadrado} \ee
Thus, the Eq. (\ref{master-eq-iii}) transforms into
\be \frac{1}{2}f\dot\psi + \left(\dot f + \lambda f^2 -\frac{3}{2}\right)\psi =
-\frac{3}{\xi}\left(1-k^2\sin^2\tau\right).
\label{1orderode}
\ee
With the aid of these two transformations, the second order differential equation
(\ref{master-eq-iii}) can be reduced to a system of first order differential equations, namely, to
a non--linear first order differential equation (general Riccati equation) for
$f(\tau)$ and a linear one for the new function $\psi(\tau)$:
\begin{eqnarray}
\dot f + \lambda f^2 -\frac{3}{2} &=& \frac{1}{2}f H(\tau), \label{eqnf} \\
\dot\psi + H(\tau) \psi &=& -\frac{3\left(1-k^2\sin^2\tau\right)}{\xi f}, \label{eqnpsi}
\end{eqnarray}
where $H(\tau)$ is a function of $\tau$. Solving this system is more easy than solving the
original differential equation (\ref{master-eq-iii}) if we make a suitable choice of the function
$H(\tau)$.

By setting $H(\tau)=2c_1/f$, with $c_1=\,$const. in the general Riccati equation (\ref{eqnf}) (this is equivalent to setting the prefactor of $\psi$ to a constant in Eq. (\ref{1orderode})) we get
the following solution for $f$:
\be
f = -\sqrt{\frac{l}{\lambda}}\tan\left[\sqrt{l\lambda}\left(\tau-\tau_0\right)\right],
\label{f}
\ee
where $2l=-(2c_1+3)$ and $\tau_0$ is an arbitrary phase. We further substitute this solution into
the linear equation (\ref{eqnpsi}) for $\psi(\tau)$  and proceed to solve it. The general solution
for this equation with arbitrary $l$ possesses a lengthy expression and is given in terms of
products of several hypergeometric and exponential functions combined with a sine at certain
power. For the sake of simplicity we shall restrict to the case in which $l=1/\lambda$, getting the
following solution:
\be
\psi = \frac{6\lambda\left[4+3\lambda-(2+3\lambda)k^2\sin^2\left(\tau-\tau_0\right)\right]}
{(4+\lambda)(2+\lambda)}+c_2\sin^{-(2+3\lambda)}\left(\tau-\tau_0\right),
\label{psig}
\ee
where $c_2$ is an arbitrary constant. Once we have a solution for $\psi$ we can get back to the
original field $\varphi=\sqrt{\psi}$ according to (\ref{phicuadrado}). This solution will be subject to
the relation (\ref{phidot}) which implies that $c_2=0$, leading to:
\be
\varphi =\sqrt{ \frac{6\lambda\left[4+3\lambda-(2+3\lambda)k^2\sin^2\left(\tau-\tau_0\right)\right]}
{(4+\lambda)(2+\lambda)}}.
\label{phi6.3.2}
\ee
In this case the $L>0$ and real $\varphi$ conditions translate into the following restrictions for $\lambda$ and $k$:
\be
a)  \quad 0<\frac{3\lambda\xi\left[(4+3\lambda)\!-\!(2+3\lambda)k^2\right]}{(4+\lambda)(2+\lambda)}<1 \quad \mbox{\rm if} \quad
\frac{\lambda(2+3\lambda)}{(4+\lambda)(2+\lambda)}<0,
\label{l3.2a}
\ee
\be
b)  \quad 0<\frac{3\lambda\xi(4+3\lambda)}{(4+\lambda)(2+\lambda)}<1 \qquad \mbox{\rm whenever} \qquad
\frac{\lambda(2+3\lambda)}{(4+\lambda)(2+\lambda)}>0,
\label{l3.2b}
\ee
where, in principle, $\lambda$ can adopt negative values.

By making use of the formula (\ref{masaplanck}), the 4D effective Planck mass for the solution (\ref{phi6.3.2}) adopts the form
\be
M_{Pl}^2 = \frac{a_{0}^3}{b\kappa}\left[1-\frac{k^2}{3}+
\lambda\,\xi\;\frac{(2+3\lambda)\,k^2-3(4+3\lambda)}{(4+\lambda)(2+\lambda)}\right].
\label{Mpl3.2}
\ee

\begin{figure}[htb]
\begin{center}
\includegraphics[width=8cm]{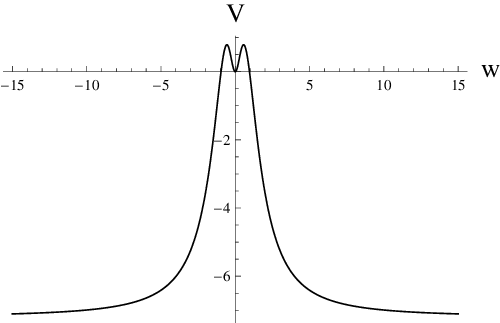}
\end{center}
\caption{$V$ associated to solution (\ref{cos}). We set  $\xi=1/3$, $a_0=1$, $b=1$, $\epsilon=1/5$ and  $\kappa=1/2$.}
\label{V_generalsol2}
\end{figure}

A further relation between $k$ and $\lambda$ simplifies the solution for
$\varphi$. For instance, when $k^2=\frac{4+3\lambda}{2+3\lambda}$, we get a sine or cosine
function depending on the phase constant $\tau_0$. It should be pointed out that both the original
differential equation (\ref{master-eq-iii}) and (\ref{eqnpsi}) are invariant under a suitable
simultaneous rescaling of the field $\varphi$ and the parameter $\xi$. This fact can be used
to arrive at the following particular solutions for the field $\varphi$:
\begin{eqnarray}
&a)& \quad \varphi=\sqrt{\frac{3}{\xi|\lambda|}}\sin(\arctan v), \qquad k^2=1-\frac{5}{2|\lambda|}\ \ \quad
\mbox{with} \quad \lambda<-5/2, \label{sin}
\\
&b)& \quad \varphi=\sqrt{\frac{6}{5\xi}}\cos(\arctan v), \qquad \ \ k^2=1+\frac{2\lambda}{5} \qquad
\mbox{with} \quad \lambda>-5/2. \label{cos}
\end{eqnarray}
In Fig. \ref{V_generalsol2} we show the profile of the $V$ for the solution (\ref{cos})\footnote{  For solution  (\ref{sin}) of the scalar field the qualitative behavior of $V$ is similar.}. Again, it is 
constant and negative at $w \rightarrow\infty$.

The 4D Planck mass for solution (\ref{sin}) reads
\be
M_{Pl}^2 = \frac{6a_{0}^3}{5b\kappa}\left(1-\frac{4}{9}\,k^2\right),
\label{Mpl3.2a}
\ee
whereas for solution (\ref{cos}) is
\be
M_{Pl}^2 = \frac{3a_{0}^3}{5b\kappa}\left(1-\frac{5}{9}\,k^2\right).
\label{Mpl3.2c}
\ee

There is another solution of the same type that is valid for the special value $\lambda=-11/8$, which implies a concrete proportional relation between $\xi$ and $\kappa$ according to (\ref{lambda}):
\begin{eqnarray}
c) \quad \varphi&=&B\cos(2\arctan v)+C, 
\label{cos2t}
\end{eqnarray}
where the following constants
\begin{eqnarray}
B_{\pm}&=&\pm\sqrt{\frac{3}{154\xi}\left[7\left(2-k^2\right)\pm
2\sqrt{4k^4+49\left(1-k^2\right)}\right]},\nonumber
\\
C_{\pm}&=&\pm\frac{\sqrt{33}\,k^2}{\sqrt{14\,\xi\left[7\left(2-k^2\right)\pm
2\sqrt{4k^4+49\left(1-k^2\right)}\right]}},\nonumber
\end{eqnarray}
must have the same sign (provided that the radicand is positive) since one can show that both
parameters $\xi$ and $k$ can be expressed as
\begin{eqnarray}
\xi&=&\frac{6}{11B^2+14BC+3C^2},\nonumber
\\
k^2&=&\frac{28BC}{11B^2+14BC+3C^2},\nonumber
\end{eqnarray}
implying that the left hand sides of these equalities must be positive.

%
%

One can look for more complex solutions than the ones displayed here. However, the relations that
define the involved integration constants become more and more lengthy and difficult their physical interpretation or viability.

From the above obtained exact solutions of this section, the field configurations (\ref{solnw=-1inx}), (\ref{sin}) and (\ref{cos}) meet the symmetry conditions (\ref{mirror-vphi})-(\ref{mirror''}) and lead to viable 4D effective theories that reproduce the correct gravitational couplings of our world.

\section{Gravitational Fluctuations}\label{sec-fluct}

In order to study the localization properties of gravity within the framework of braneworlds, a rigorous analysis of gravitational
fluctuations is needed.

In 4D standard cosmology, the fluctuations of the geometry
can be classified into scalar, vector and tensor modes with respect to the three--dimensional
rotation group $SO(3)$ \cite{bardeen,stewart}. This fact makes more feasible the study of the metric
fluctuations because at the linear level the dynamical equations of the scalar, vector and tensor
modes are decoupled.

Within the braneworld models considered in our work, where the 4D geometry is Poincar\'e invariant,
the fluctuations of the metric may be also classified into scalar, vector and tensor modes with
respect to the transformations of the $SO(3,1)$ symmetry group (see \cite{padilla} for details).

Let us consider the fluctuations of both the metric and the scalar field around the gravitational
background specified in (\ref{line-e}) and the field equations (\ref{eequation1})-(\ref{eequation3}). In other words, the
perturbed geometry has the following form
\begin{equation}
ds_{p}^2 =\left[a^2(w) \eta_{A B}+H_{AB}(x,w)\right]dx^{A}dx^{B}, \quad \mbox{where} \quad x\equiv \{x^{\mu}\}
\label{elinep},
\end{equation}
while the fluctuation of the scalar field is
\begin{equation}\nonumber
\varphi_{p}=\varphi(w) + \chi(x,w).
\end{equation}
The functions $H_{A B}(x,w)$ and $\chi(x,w)$ are the metric and the field fluctuations, respectively. The
index $p$ denotes perturbed quantities.

By taking into account the 4D Poincar\'e symmetry of our background metric
(\ref{line-e}), the  fluctuations can be written as
\begin{equation}\nonumber
H_{AB} = H^{(S)}_{AB} +H^{(V)}_{AB}     + H^{(T)}_{AB},
\end{equation}
where
\bea
H^{(S)}_{AB}\!\!&=&\!\!a^2(w)
\left( \begin{array}{cc} 2\left(\eta_{\mu \nu} \psi
+ \partial_{\mu}\partial_{\nu} E\right) & \partial_{\mu} C\\ \partial_{\mu} C  &2 \zeta \\
\end{array}\right),\label{fluc_escalar} \\
H^{(V)}_{AB}\!\!&=&\!\!a^2(w)
\left( \begin{array}{cc}
(\partial_{\mu} f_{\nu} +\partial_{\nu} f_{\mu})
 & D_{\mu} \\ D_{\mu}   &0 \\
\end{array}\right),\label{fluc_vectorial} \\
H^{(T)}_{AB}\!\!&=&\!\!a^2(w)
\left( \begin{array}{cc} 2 h_{\mu\nu}
 & 0\\ 0 &0 \\
\end{array}\right).\label{fluc_tensorial}
\eea The upper indices $S$, $V$ and $T$ denote the scalar, vector and tensor parts of the
fluctuations, respectively. The tensor $h_{\mu \nu}(x,w)$ is transverse and traceless with respect to
the 4D Minkowski metric $\eta_{\mu \nu}$, in other words \be
\label{tensor_cond_norma}
 h_{\mu}^{\mu} = 0,\,\,\,\,\,\,\,\,\,\,\,
\partial_{\nu} h_{\mu}^{\nu}= 0.
\ee
Also, the vectors $f_{\mu}(x,w)$ and  $D_{\mu}(x,w)$ are divergence free
\be  \label{vector_cond_norma}
\partial^{\mu}f_{\mu} = 0,\,\,\,\,\,\,\,\,\,\,\,
\partial^{\mu}D_{\mu} = 0.
\ee The four remaining functions $\psi(x,w)$, $E(x,w)$, $C(x,w)$ an $\zeta(x,w)$ are scalars with respect to
4D Poincar\'e transformations.

The relations (\ref{fluc_escalar})--(\ref{vector_cond_norma}) tell us that, apparently, we only
have $15$ independent degrees of freedom of the metric fluctuations. Moreover, the covariance  of
our setup implies that the gravitational perturbation theory has some unphysical gauge degrees of freedom
\cite{bardeen,stewart}. In our case we can make use of this gauge freedom completely by fixing $5$ of the above $15$ degrees of freedom.

A simple choice that completely fixes the above mentioned gauge freedom is the longitudinal gauge given by:
\begin{equation}
E = 0, ~~~~ C =0, ~~~~f_{\mu} = 0.
\label{long_norm}
\end{equation}
Thus, we finally have only $10$ independent degrees of freedom.

\subsection{Tensor modes}

As quoted in \cite{CQG29} the equation for  the evolution of the tensor fluctuation modes is
\begin{equation}
\Psi_{\mu \nu}'' - \frac{({\sqrt{s}})''}{\sqrt{s}} \Psi_{\mu \nu} -
\frac{r}{s} \Box^{\eta}\Psi_{\mu
\nu}=0,\label{graviton1}
\end{equation}
where $\Psi_{\mu \nu}
=\sqrt{s(w)} h_{\mu \nu}$ with $s(w) = a^{3} q$. In addition, $r(w) =a^{3}\biggl( q + \frac{(q-L)'}{2{\cal H}}\biggr)$ and $\Box^{\eta}$ denotes the d'Alembertian with respect to the metric $\eta$.

In order to study the mass spectrum of these fluctuations let us consider the following separation
of variables $\Psi_{\mu \nu}(x,w)= \vartheta(w)\epsilon_{\mu \nu}(x)$. In terms of this new variables
the equation (\ref{graviton1}) splits into
\begin{eqnarray}
\Box^{\eta} \epsilon_{\mu \nu}^{m}+m^2 \epsilon_{\mu \nu}^{m}&=&0,\label{graviton4d}\\
\vartheta_{m}'' - \frac{({\sqrt{s}})''}{\sqrt{s}} \vartheta_{m} +m^2
\frac{r}{s}\vartheta_{m}&=&0,\label{graviton5d}
\end{eqnarray}
where  $\epsilon_{\mu \nu}^{m}(x)$ is a 4D tensor mode with mass $m$; on the other
hand, $\vartheta_{m}(w)$ is the 5D profile of the field $\Psi_{\mu \nu}$ and
characterizes its localization properties. The Eq. (\ref{graviton5d}) can be interpreted as a
Sturm--Liouville eigenvalue problem with the associated  norm:
\begin{equation}
\langle \vartheta |\vartheta \rangle=\int_{-\infty} ^{\infty}
\frac{r}{s}\,\vartheta^2 \,dw. \label{norma_tensorial}
\end{equation}
The zero mode $\epsilon_{\mu \nu}^{0}(x)$  is the massless 4D graviton. This mode
is localized on the brane if $\langle \vartheta_0 |\vartheta_0 \rangle=\int_{-\infty} ^{\infty}
\frac{r}{s}\,\vartheta_0^2 \,dw $ is finite. It is not difficult to show from (\ref{graviton5d}) that
$\vartheta_0=\sqrt{s}$ when $m=0$, then, the localization condition for the massless  graviton is
\begin{eqnarray}
&\langle \vartheta_0 |\vartheta_0 \rangle=\int_{-\infty}^{\infty} a^3\,q \,dw
-4 \,\epsilon \,[a']_{-\infty}^{\infty} +8\,\epsilon
\int_{-\infty}^{\infty}
\frac{a'^2}{a}\,dw.\label{norma_cero_tensorial}
\end{eqnarray}
As one can observe the above expression is quite similar
to (\ref{masaplanck}). Therefore, {\it for the geometry described in (\ref{warp-f}), if the 4D massless graviton is localized on the brane, then the 4D Planck mass is finite}.

Let us investigate the localization properties of the massless graviton in the backgrounds studied
in Section \ref{exact-sol}.  By substituting the expression of the warp factor into
(\ref{norma_cero_tensorial}) we obtain the following result
\begin{eqnarray}
&\langle \vartheta_0
|\vartheta_0 \rangle_{GB}=\frac{2 a_{0}^3}{b}\left(1+\frac{k^2}{3}\right)
\label{norm_tensor_GB}.
\end{eqnarray}
The above expression tells us that the 4D graviton is localized on the brane. This result generalizes the particular situation considered in \cite{mg00} where $\xi = 0$, $\epsilon \neq 0$ and $k=1$ for an arbitrary value of $k$, showing that the zero mass graviton is also localized on the brane.

In the following we shall consider the opposite case where $\xi \neq 0$ and $\epsilon =0$. The
normalization condition depends of the behavior of the integrand $a^{3} L(\varphi)$. With regard
to the study of gravity localization, the backgrounds defined in (\ref{sol-ii-1'}) and
(\ref{sol-ii-2'}) are similar because both have a regular and finite non--minimal coupling
function $L(\varphi)$. Therefore, the convergence of (\ref{norma_cero_tensorial}) is completely
determined by the $a^3$ behavior when $w \rightarrow \infty$. For the warp factor (\ref{warp-f})
$a^3 \sim 1/|w|^3$ at infinity along the fifth dimension. This fact implies again that the massless tensor mode is
localized on the brane, since the third term in (\ref{norma_cero_tensorial}) is finite.

Finally, let us study the general case ($\xi \neq 0$ and  $\epsilon \neq  0$). Some exact
solutions for this general situation are shown in (\ref{solnw=-1inx}), (\ref{sin}), (\ref{cos}) and (\ref{cos2t}).
The norm for the zero mass mode takes the following form
$$
\langle \vartheta_0 |\vartheta_0 \rangle=\int_{-\infty}^{\infty} a^3\,(L-1) \,dw + \langle
\vartheta_0 |\vartheta_0 \rangle_{GB},
$$
where $\langle \vartheta_0 |\vartheta_0 \rangle_{GB}$ is
the norm written in (\ref{norm_tensor_GB}). On the other hand, the function $L$ associated to
(\ref{solnw=-1inx}), (\ref{sin}), (\ref{cos}) and (\ref{cos2t})  is  finite along the extra dimension, thus, like
in the previous cases the normalization condition depends on the behavior of $a^3$ at infinity.
Hence,  we conclude that the zero tensor mode is normalized. In summary, {\it all our background solutions
recover the standard 4D  graviton on the brane for the case when both the non--minimal coupling and the Gauss--Bonnet term are 
present in the model.}

\subsection{Vector modes}\label{sec_vm}

In contrast with the tensor sector, the simultaneous presence of the Gauss--Bonnet term and the
non--minimal coupling interaction makes much more difficult the study of vector modes. Thus, in
this work we will study each case separately and will leave the general case for a future
investigation.

\subsubsection{Vector modes with the Gauss--Bonnet term only.}

In this subSection we analyze the  case where the non-minimal effects are negligible ($\xi=0$ and
$\epsilon \neq 0$). Using (\ref{elinep}), (\ref{fluc_vectorial}), (\ref{vector_cond_norma}) and
substituting them in (\ref{ee}) we obtain the equations for the vector modes on the longitudinal
gauge
\begin{eqnarray}
&&(D^{\nu})'+\left(\frac{q'}{q} +3{\cal H} \right)D^{\nu}=0,
\label{ecua_fluc_vec_munu0}\\
&&\Box^{\eta} D_{\mu}=0,\label{ecua_fluc_vec_wmu0}
\end{eqnarray}
where the first relation is a constraint and defines the profile of $D_{\mu}$ along the extra
dimension. Furthermore, since there is no mass term in the above second equation, it shows that
there is only a massless mode, the graviphoton, with no massive vector fluctuations.

In the case of vector modes we do not have a Sturm--Liouville eigenvalue problem, in consequence,  the issue of defining the norm is more involved than for the tensor sector. Therefore, it is more appropriate to use the perturbed version of the action (\ref{action})
up to second order with respect to the vector fluctuations \cite{mg1,mg00}. This perturbed action
can be written as follows
\begin{equation}
\delta^{(2)} S_{V} = \int d^{4} x d w
\frac{1}{2}\left( \eta^{\alpha\beta}
\partial_{\alpha} {\cal D}^{\mu} \partial_{\beta} {\cal D}_{\mu} \right),
\label{canvec}
\end{equation}
where ${\cal D}_{\mu} = a^{3/2} \sqrt{q} D_{\mu}$ is called {\it canonical normal mode}. The zero
mode associated to (\ref{ecua_fluc_vec_wmu0}) is localized on the brane if $\delta^{(2)} S_{V}$ is
finite. By making a suitable variable separation
\begin{equation}
{\cal D}^{\mu} = \frac{v^{\mu}(x)}{a^{3/2}(w) \sqrt{q(w)}},\nonumber
\end{equation}
the zero mass mode takes the form
\begin{equation}
\Box^{\eta} v^{\mu}(x)=0, \nonumber
\end{equation}
where $v^{\mu}(x)$ is the 4D part of the vector sector of fluctuations.

The norm of this mode reads
\begin{equation}\label{norma_vectorial}
\langle {\cal D}^{\mu}|{\cal D}_{\mu}\rangle=
\int_{-\infty}^{\infty} \frac{dw}{a^{3} q}.
\end{equation}
When studying the localization properties of the zero mass tensor mode, we learned that it is
necessary to have a convergent integral of $a^3 q$. Furthermore, the norm of the massless vector
mode has an opposite behavior compared to the tensor mode one. Therefore, if one wishes the
4D graviton to be localized on the brane, the vector sector (represented by its
zero mass mode) will necessarily not be localized on it.

\subsubsection{Vector modes with the non--minimal coupling  only.}

In this subSection we shall study the situation when $\epsilon = 0$ and $\xi \neq 0$. One way to
obtain the mass spectrum of the vector fluctuations consists in passing from the action
(\ref{action}) (with $\epsilon = 0$) to the Einstein frame and then, perform the perturbation
analysis. In order to do that we apply a conformal transformation as follows
$$
\overline{g}_{A B}=L^{2/3}g_{A B}.
$$
In the new frame the background action can be written as
\begin{eqnarray}
&S \mapsto S_{EF} = \int_{M_{5}} d^5 x \sqrt{|\overline{g}|} \biggl\{- \overline{R}
+ \frac{1}{2} (\overline{\nabla} \sigma)^2 -
\overline{V}(\sigma)  \biggr\}, \label{actioneinsteinf}
\end{eqnarray}
where all quantities with an overline are associated to the  metric $\overline{g}_{A B}$, $\sigma$
is the new scalar field and $\overline{V}$ is its self--interaction potential. The relationship
between the old and new variables is
\begin{equation}
d\sigma=\sqrt{\frac{1}{L}
+\frac{8}{3}
\left(\frac{L_{\varphi}}{L}\right)^{2}} d\varphi, \label{campo_marco_eist}
\end{equation}
\begin{equation}\nonumber
\overline{V}=\frac{V}{L^{5/3}}.
\end{equation}
Moreover, the general fluctuations on the Einstein frame are related with the old quantities as
follows
\begin{equation}
\overline{H}_{AB}=\frac{2 a^{2}}{3 L^{1/3}}
\chi L_{\varphi} \eta_{A B}
+ L^{2/3} H_{A B}.\label{fluctuaciones_conforme}
\end{equation}
The first term in the above expression does not contribute to the vector sector since it belongs
to the scalar sector, then it is easy to obtain that $\overline{D}_{\mu}=D_{\mu}.$ Thus, the
equations for the vector modes are
\begin{eqnarray}
&&(\overline{D}^{\nu})' +3\overline{{\cal H}}\,\overline{D}^{\nu}=0,
\nonumber \\
&&\Box^{\eta} \overline{D}_{\mu}=0,\nonumber
\end{eqnarray}
where $\overline{{\cal H}}=\overline{a}'/ \overline{a}$ and $\overline{a}=a L^{1/3}$.

Similarly to the previous case, let us define a new vector variable $\overline{{\cal D}}_{\mu} =
\overline{a}^{3/2} \overline{D}_{\mu}$. Therefore, the norm of the  massless mode takes the form
\begin{equation}\nonumber
\langle \overline{{\cal D}}^{\mu}|\overline{{\cal D}}_{\mu}\rangle=
\int_{-\infty}^{\infty} \frac{dw}{\overline{a}^{3}}=
\int_{-\infty}^{\infty} \frac{dw}{a^{3}L}.
\end{equation}
The localization properties of this case are similar to those of the previous one. The
localization of the massless tensor mode implies the delocalization of the vector sector of
fluctuations. In other words, {\it if the effective Planck mass is finite (or the 4D massless graviton is localized on the brane), then the vector sector is not confined to the brane.}

In principle, it still can have observable effects in the 4D phenomenology since its projection to the brane can be non--vanishing. Notwithstanding, this delocalization phenomenon of the vector modes implies that they have little influence in the 4D low energy physics, at 
least less influence than the tensor sector.

\subsection{Scalar modes}\label{sec_scalarm}

In the longitudinal gauge the scalar fluctuations of the geometry can be expressed as
$$
H^{(S)}_{AB}=2a^2(w)
\left( \begin{array}{cc} \eta_{\mu \nu} \psi
 & 0\\ 0  & \zeta \\
\end{array}\right).
$$
Furthermore, the matter provides an extra degree of freedom $\chi$ to the scalar sector. Hence, it
follows that we have three independent scalar degrees of freedom.  One can obtain the dynamical
equations of the scalar fluctuations by perturbing the Einstein and Klein--Gordon equations with
respect to the scalar sector. These equations adopt the following form
\begin{eqnarray}
&& \!\!\!\!4 q \psi''
+ 4\psi' \biggl[q+\frac{q'}{{\cal H}}+
\frac{\varphi' L_{\varphi}}{3}\biggr]  +8 \zeta \biggl[ {\cal H}'L +\frac{\varphi'^{2}}{8}
 - \frac{ 4 \epsilon {\cal H}^2}{a^2} (
2 {\cal H}' - {\cal H}^2)  + \frac{(\varphi' L_{\varphi})'}{3} \biggr]
\nonumber\\
&&\!\!\!\! + 4 \zeta' \biggl[
\frac{\varphi' L_{\varphi}}{3} +{\cal H} q\biggr]+
q \Box^{\eta} \zeta + \frac{8 \epsilon}{a^{2}}({\cal H}' -
 {\cal H}^{2})\Box^{\eta}\psi
+ \frac{1}{3} \frac{d V}{d \varphi} a^2 \chi \label{fescalarww}\\
&&\!\!\!\!+ \varphi' \chi' +4 {\cal H}' \chi L_{\varphi}+ \frac{1}{3}\biggl[ 4 (\chi L_{\varphi})''
-\Box^{\eta} (\chi L_{\varphi})  \biggr]=0.\nonumber
\end{eqnarray}
\begin{eqnarray}
&&  \!\!\!\!\!q \psi''
+ \psi' \biggl[7{\cal H} L
-\frac{4 \epsilon {\cal H}}{a^{2}} \left( 2 {\cal H}'+ 5 {\cal H}^{2}\right)
+\frac{7 \varphi' L_{\varphi}}{3}\biggr] +  \zeta' \biggl[
\frac{\varphi' L_{\varphi}}{3} +{\cal H} q\biggr]\nonumber\\
&& \!\!\!\!\!+2\zeta \biggl[ (3{\cal H}^{2}\!+\! {\cal H}')L
\! -\! \frac{ 8 \epsilon {\cal H}^2}{a^2} (
{\cal H}' \!+\! {\cal H}^2) +2{\cal H} \varphi' L_{\varphi}
\!+\! \frac{(\varphi' L_{\varphi})'}{3} \biggr] +2 {\cal H} (\chi L_{\varphi})'
\label{fescalarmunu}\\
&&\!\!\!\!\!
+ \frac{1}{3} \frac{d V}{d \varphi} a^2 \chi -
q \Box^{\eta} \psi \!+\!( {\cal H}'\!+\!3{\cal H}^{2}) \chi L_{\varphi}\!+\! \frac{1}{3}\biggl[  (\chi L_{\varphi})''
\!-\!\Box^{\eta} (\chi L_{\varphi})  \biggr]=0,\nonumber
\end{eqnarray}
\begin{eqnarray}
&&q \zeta - 2 \psi \biggl[ L - \frac{4 \epsilon {\cal H}'}{a^2} \biggr]-\chi L_{\varphi}=0.
\label{fescalarnomunu}\\
 &&\!\frac{\varphi' \chi}{2} \!+\! 3 q (\psi' \!+\! {\cal H} \zeta) \!+\!
(\chi L_{\varphi})' \!-\!{\cal H}\chi L_{\varphi} \!+\!L' \zeta =0.
\label{fescalarmuw}
\end{eqnarray}
\begin{eqnarray}
&&\chi'' + 3{\cal H} \chi' - \Box^{\eta} \chi
+ \varphi'[4 \psi' + \zeta'] + 2\zeta ( \varphi'' + 3 {\cal H} \varphi' )\nonumber\\
&&- \frac{\partial^2 V}{\partial\varphi^2}a^2 \chi
-\frac{1}{2}\biggl\{  2(2{\cal H}' + 3 {\cal H}^2)
\chi L_{\varphi \varphi} +L_{\varphi}\left[\Box^{\eta}\zeta \right. \label{e_kg_sector_escalar}\\
 &&\left. -
3\Box^{\eta}\psi +4 \zeta(2{\cal H}' + 3 {\cal H}^2)+4(\psi''
+{\cal H} [4\psi' +\zeta'] )\right]  \biggr\}=0.\nonumber
\end{eqnarray}
Eqs. (\ref{fescalarww})-(\ref{fescalarmunu}) come from the perturbed Einstein equations, while
the expressions (\ref{fescalarnomunu}) and (\ref{fescalarmuw}) are constraint equations. Finally, Eq. (\ref{e_kg_sector_escalar}) represents the fluctuated Klein--Gordon equation. Then,
there is only one independent scalar degree of freedom. As in the vector sector, in this case we
will study separately the effects of the Gauss--Bonnet term and the non--minimal coupling on the
scalar modes.

\subsubsection{Scalar modes with the Gauss--Bonnet term only.}

Although some aspects of this case were studied in \cite{mg00}, it is helpful to consider some of
its details. The master equation of the system can be obtained by using the constraints
(\ref{fescalarnomunu}), (\ref{fescalarmuw}) and the dynamical equations (\ref{fescalarww}),
(\ref{fescalarmunu}) and (\ref{e_kg_sector_escalar}):
\begin{equation}
\Phi'' - z \biggl(\frac{1}{z}\biggr)'' \Phi
- \biggl( 1 + \frac{ q'}{{\cal H} q} \biggr)  \Box^{\eta}\Phi =0,
\label{ph}
\end{equation}
where $\Phi = \frac{ a^{3/2} q}{\varphi'}\psi$ and $z = \frac{a^{3/2} \varphi'}{{\cal H}}$. In
order to study the mass spectrum of $\Phi$ let us assume that
$$
\Box^{\eta}\Phi=-m^{2}\Phi.
$$
Thus, the Eq. (\ref{ph}) can be written as
\begin{equation}
\Phi'' - z \biggl(\frac{1}{z}\biggr)'' \Phi
+\biggl( 1 + \frac{ q'}{{\cal H} q} \biggr)  m^{2}\Phi =0.
\label{ph1}
\end{equation}
The associated norm for the above eigenvalue problem is
\begin{equation}\label{norma_escalar}
\langle \Phi | \Phi \rangle=\int_{-\infty}^{\infty}
\biggl( 1 + \frac{ q'}{{\cal H} q} \biggr) \Phi^{2} dw.
\end{equation}
The scalar  massless mode is localized on the brane if its norm is finite. In other words
\be\label{norma_cero_escalar} \langle \Phi_{0} | \Phi_{0} \rangle=\int_{-\infty}^{\infty} \biggl(
1 + \frac{ q'}{{\cal H} q} \biggr) \Phi_{0}^{2} dw < \infty, \ee where $\Phi_{0}=\frac{1}{z}$. Let
us apply this general analysis to the background solution (\ref{sol-i-1}). It is not hard to show
that when $w \rightarrow \infty$
\begin{eqnarray}
 \biggl( 1 + \frac{ q'}{{\cal H} q} \biggr) \Phi_{0}^{2}\sim \left\{ \begin{array}{ll}
 |w|^7, & \qquad\mbox{when}\qquad k=1 \\
 & \\
 |w|^5, & \qquad\mbox{when} \qquad k \neq 1\\
\end{array}
\right.
\end{eqnarray}
Therefore, in both cases the massless scalar mode is not localized on the brane.

\subsubsection{Scalar modes with the non--minimal coupling only.}

Similarly to the vector modes case, the analysis of the scalar sector without the Gauss--Bonnet
term is easier in the Einstein frame. In this case the metric fluctuations can be expressed as
follows
\begin{eqnarray}
\overline{H}^{(S)}_{AB}\!\!&=&\!\!2 \overline{a}^2
\left( \begin{array}{cc} \eta_{\mu \nu} \overline{\psi}
& 0\\ 0 &\overline{\zeta}, \\
\end{array}\right),\label{fluc_escalar_conforme}
\end{eqnarray}
where $ \overline{\psi}=\psi +\frac{\chi L_{\varphi}}{3 L}$ and $\overline{\zeta}=\zeta
-\frac{\chi L_{\varphi}}{3 L}$. On the other hand, the scalar field fluctuation in the
Einstein frame $\overline{\chi}$ is related to $\chi$ as follows
$$\overline{\chi}=\sqrt{\frac{1}{L}
+\frac{8}{3}
\left(\frac{L_{\varphi}}{L}\right)^{2}}\,\, \chi.
$$
By considering the above expression and the Eqs.
(\ref{fescalarww})--(\ref{e_kg_sector_escalar}) when $\epsilon=0$, we obtain the following equation
for the scalar sector of fluctuations
\begin{equation}
\overline{\Phi}'' - \overline{z} \biggl(\frac{1}{\overline{z}}\biggr)'' \overline{\Phi}
+ m^{2}\overline{\Phi} =0,
\label{ph_conforme}
\end{equation}
where $\overline{\Phi} = \frac{ \overline{a}^{3/2}}{\sigma'}\overline{\Psi}$ and
$\overline{z}=\frac{\overline{a}^{3/2} \sigma'}{{\cal \overline{H}}}$. Thus, the norm of zero mode
reads
\be\label{norma_cero_escalar_conforme} \langle \overline{\Phi}_{0} | \overline{\Phi}_{0}
\rangle=\int_{-\infty}^{\infty}
 \overline{\Phi}_{0}^{2} dw=\int_{-\infty}^{\infty}
 \frac{1}{\overline{z}^{2}} dw.
\ee The localization properties of the solutions (\ref{sol-ii-1'}) and (\ref{sol-ii-2'}) are
similar in the sense that their massless mode has the same behavior at infinity, i.e.
$$
\frac{1}{\overline{z}^{2}} \sim |w|^5, \qquad \mbox{when}\qquad w\rightarrow \infty.
$$
 The above result tells us that the zero mass scalar mode is not localized on the brane.

Therefore, the results of the subsections \ref{sec_vm} and  \ref{sec_scalarm} tell us that both
the vector and the scalar fluctuations modes behave in a quite similar way when regarding their localization
properties, both are delocalized from the brane if 4D gravity is localized on it or, equivalently, if the 4D planck mass is finite.

\section{Conclusions}

We present a scenario with  a thick braneworld model builded by a scalar field  non--minimally coupled to the Einstein-Hilbert term. Furthermore, there is a Gauss-Bonnet term in the bulk.
We obtain the effective 4D Planck mass coming from the dimensional
reduction of the set--up (\ref{action}) and study its properties. As a consequence  of the non--minimal
coupling between the scalar matter and gravity, $M_{\rm Pl}$ depends explicitly on the profile of the bulk scalar field $\vphi$. Therefore, in order for us to be able to interpret our physical results in closed form we
need to construct exact solutions for our braneworld configuration.

Despite the highly non--linear
nature of the relevant differential equation for the scalar field, we were able to obtain
several exact particular solutions, contrary to previous studies that implement approximate
solutions and/or particular cases that consider just one of the two effects.

In order to elucidate
the physical effects of the inclusion of the non--minimal coupling and of the Gauss--Bonnet term,
we considered three cases in which we switched on/off the corresponding parameters. In all of these cases we were able to construct exact solutions which render an effective Planck mass positive and finite by imposing a restriction on the non-minimal coupling function $L(\varphi)$ for certain values of the parameter space of these solutions. However, there are also exact field configurations for which the 4D Planck mass is negative, yielding 
braneworld models that cannot physically describe the gravitational interactions of our world.

We also studied the
whole set of gravitational fluctuations--classified into scalar, vector and tensor modes with
respect to the 4D Poincar\'e symmetry group -- for our braneworld model. For all the considered
backgrounds the massless zero mode of the tensor fluctuations is localized on the brane. This fact
allows us to recover 4D gravity in our world. In contrast to this, the massless 
scalar and vector modes are delocalized.

Although the scalar and vector sectors have a non--zero projection on the brane, due 
to the delocalization phenomenon it is not strange that for our scenario they do not 
have sensible observable effects at low energies. As we commented above, this last 
assertion is reinforced by the fact that in a simple scenario, the corrections to 
Newton's law coming from the scalar modes are more suppressed than the corrections 
coming from the tensor modes. Therefore, the above result suggests that in gravitational 
experiments it is more probable to detect first the effects 
of the tensor sector rather than the effects of the scalar and vector ones.
It is noteworthy that the delocalization  
phenomenon on thick branes also appears when  the non--minimal coupling and the 
Gauss--Bonnet are not present in our scenario \cite{mg1}.
Then, the above observation suggests us that in thick braneworlds with 4D Poincar\'e invariant geometries
the decoupling of the vector and scalar sectors
of the 4D phenomenology is a robust effect of this kind of scenarios.

As a further step it will be interesting to study the stability of the field
configurations within the framework of the present braneworld model when both non--linear
effects are switched on. It is not completely clear if the delocalization of the vector and scalar sectors
is maintained in other class of solutions not studied here, this seems to be a very suggestive direction.
Another interesting topic is to consider a non--minimal coupling of the
bulk scalar field not only with the Einstein-Hilbert piece of the Lagrangian but, also, with the
Gauss-Bonnet term. Investigation of these issues is left for future work.

\section{Acknowledgements}

AHA, DMM and IQ are grateful to U. Nucamendi, A. Raya and L. Ure\~{n}a for useful discussions. AHA
thanks the staff of the ICF, UNAM  and MCTP, UNACH for hospitality. 
GG, AHA, DMM and RR gratefully acknowledge support
from \textquotedblleft Programa de Apoyo a Proyectos de Investigaci\'on e Innovaci\'on
Tecnol\'ogica\textquotedblright\, (PAPIIT) UNAM, IN103413-3, {\it Teor\'ias de Kaluza-Klein,
inflaci\'on y perturbaciones gravitacionales.} DMM acknowledges a Postdoctoral grant from
DGAPA-UNAM. GG,  AHA and DMM thank SNI for support.
IQ acknowledges support from "Programa PRO-SNI, Universidad de Guadalajara" under grant No 146912.
RR is grateful to Conselho Nacional de
Desenvolvimento Cient\'{\i}fico e Tecnol\'ogico (CNPq) grants 476580/2010-2 and 480482/2012-8 for
financial support.

\end{document}